\documentclass[12pt]{iopart}

\usepackage{graphicx}
\begin{document}

\title{How do signs organize in directed signed social networks?}

\author{Long Guo,  Fujuan Gao}
\address{School of mathematics and Physics, China University of Geosciences (Wuhan),\\ Lumo Road, 430074, Wuhan, China}

\ead{guolong@cug.edu.cn (Long Guo)}

\begin{abstract}
We introduce a reshuffled approach to empirical analyze signs' organization in real directed signed social networks of Epinions and Slashdots from the global viewpoint. In the reshuffled approach, each negative link has probability $p_{rs}$ to exchange its sign with another positive link chosen randomly. Through calculating the entropies of social status ($S_{in}$ and $S_{out}$) of and mimicking opinion formation of the majority-rule model on each reshuffled signed network, we find that $S_{in}$ and $S_{out}$ reach their own minimum values as well as the magnetization $|m^{*}|$ reaches its maximum value at $p_{rs}=0$. Namely, individuals share the homogeneous properties of social status and dynamic status in real directed signed social networks. Our present work provides some interesting tools and perspective to understand the signs' organization in signed social networks.

\end{abstract}

\pacs{89.70.Cf; 89.75.Fb; 89.65.Ef}
\maketitle

\section{Introduction}
\label{intro}
Recently, complex networks have undergone a remarkable development and have emerged as an invaluable tool for describing and quantifying complex systems in physics, biology and sociology. Generally, a complex network is usually described by a graph in which vertices represent the components, such as people in social network and proteins in protein-protein interaction network, and links represent the interaction among those components. In most cases, links are all considered as positive connections, for example links in social networks indicate friendship, collaboration and sharing information\cite{sun2014}. However, many real social networks, especially online social networks (such as the EBay, Epinions and Slashdot), intrinsically involve negative links as well as positive ones, such as the enemy, disproval and distrust relationships. For example, users can tag directed relations to others indicating trust or distrust in the trust network of Epinions, and users can designate others as "friends" or "foes" in the social network of the technology blog Slashdot\cite{leskovec20101}. Those social networks can be represented in terms of signed social networks\cite{facchetti2012, esmailian2014, ciotti2015}, where a sign of link is defined as "+1" or "-1" depending on whether it expresses a positive or negative attitude from the generator of the link to the recipient\cite{leskovec20102}.

The fundamental questions are how do signs organize in real signed social networks\cite{leskovec20102} and how does the real sign organization affect the dynamics of and on signed social networks. For the first issue of signs' organization, the social balance theory was proposed by Heider from the aspect of social psychology\cite{facchetti2011}.  The social balance theory has recently attracted more attention from sociologists and physicists. Facchetti et al.\cite{facchetti2011} computed the global level of balance of online signed social networks and found that the currently available undirected networks, such as the Epinions and Slashdot, are indeed extremely balanced. Traag et al. proposed an alternative model based on the homophily process to explain the social balance and the evolution of cooperation\cite{traag2013}. Furthermore, Facchetti et al.\cite{facchetti2012} investigated the organization of frustration through exploring the low-energy landscape of near-optimal structural balance from the aspect of statistics mechanics. However, some researches shown that many signed social networks, especially the directed online social networks, are very poorly balanced\cite{leskovec20101, estrada2014}. Then, Leskovec et al.\cite{leskovec20102} developed the status theory, where a positive edge $(u, v)$ means that $u$ regards $v$ as having higher status than himself/herself while a negative edge $(u, v)$ means that $u$ regards $v$ as having lower status than himself/herself, to explain the signs' organization.

The second issue about the role of the real signs' organization in the dynamics of and on signed social networks has also been studied in the last decades\cite{summers2013, singh2014, nishi2014, fan2012, li2012, righi2014, jiang2015}. For example, Nishi and Masuda analyzed the dynamics of social balance under undirected temporal interaction. And they found that the social balance dynamics is slowed down on the temporal complete network through compared to the corresponding static complete network\cite{nishi2014}. Fan et al. \cite{fan2012} analyzed the opinion spreading based on the SIR model in homogeneous signed networks.  Li et al. extended the classic voter model to signed networks and analyzed the dynamics of influence diffusion of two opposite opinions\cite{li2012}. Righi and Tak\'{a}cs studied the Prisoner's Dilemma on signed networks where the behavior of condition player is determined by link's sign\cite{righi2014}.

 Our main goal is to evaluate the real signs' organization through calculating the information entropies of social status and the critical order parameter of opinion formation in the directed signed social networks of Epinions and Slashdot. Each negative link has the reshuffled probability $p_{rs}$ to exchange his/her sign with another positive link chosen randomly. Then we analyze the information entropies $S_{out}$ and $S_{in}$ of each reshuffled signed network. We find that individuals tend to share the homogeneous social status in the real signed social networks. Furthermore, in order to reveal the role of the real signs' organization in the opinion formation, the majority-rule model is realized on each reshuffled signed network. We find that the critical order parameter $|m^{*}|$ reaches its maximum value when $p_{rs}=0$, which means that even more individuals share the same opinion in real signed social networks. The present signs' organization enhances the consensus ability of the system, i.e., individuals have the tendency to share the homogeneous dynamic status in collective dynamics of opinion formation.

\section{\label{sec:level2}Description of signed social networks and the reshuffled approach}

We consider a directed signed social network $G=(V, L, A)$, where $V$ is the set of vertices, $L$ denotes the set of directed links, $A=\{A_{uv}\}$ describes the signed adjacency matrix with $A_{uv}\neq 0$ if and only if $(u,v)\in L$, and $A_{uv}$ is the sign of link $(u,v)$. A positive sign $A_{uv}(=+1)$ represents that $u$ tags $v$ as a friend or $u$ trusts $v$, while a negative sign $A_{uv}(=-1)$ reflects that $u$ tags $v$ as a foe or $u$ distrusts $v$. Several real signed social networks\cite{leskovec20102}, including the network of Epinions was obtained in August 12, 2003, the network of Slashdot081106 was obtained in November 6, 2008, the network of Slashdot090216 was obtained in February 16, 2009 and the network of Slashdot090221 was obtained in February 21, 2009, are considered and available at http://snap.stanford.edu. The trust network of Epinions is a product review Website with a very active user community, where users can tag their trust or distrust of the reviews of others, and the social network of the Slashdot is a technology-related network website, where a signed link means that one user likes or dislikes the comments of another user. In each network, links are inherently directed and the proportion of positive links is roughly 80\%\cite{leskovec20101}, see Table.\ref{tab.1}.

\begin{table}
\caption{Nodes and links in several signed social networks (Epinions and Slashdot)\cite{leskovec20102}. $N_{v}$ is the number of the vertices, $N_{l}$ is the number of the directed links and $p_{+}$ is the percent of sign "+" (i.e., the percent of the positive links).}
\label{tab.1}
\begin{center}
\begin{tabular}{lcrr}
 & $N_{v}$ & $N_{l}$ & $p_{+}$ \\
Epinions & 131,828 & 841,372 & 85.3\%\\
Slashdot081106 & 77,357 & 516,575 &76.7\%\\
Slashdot090216 & 81,871 & 545,671 &77.4\%\\
Slashdot090221 & 82,144 & 549,202 &77.4\%
\end{tabular}
\end{center}
\end{table}

A given signed social network with constant macroscopic quantities ($N_{v}$, $N_{l}$, $p_{+}$) can be regarded as an isolated system, which obeys the ergodic hypothesis from the viewpoint of statistic mechanics. According to the ensemble theory in statistical mechanics, each signed social network has many configurations related to signs' organization. As well known, to flip one or more signs will alter its signs' organization, and the system will experience another microscopic state. Different signs' organization describes different microscopic state. There must exist one specific microscopic state which has the same signs' organization of the real signed social network. In order to analyze the difference of the possible signs' organization and the real one, we introduce a reshuffled approach to rebuild configurations of the real signed social networks of Epinions and Slashdot through the tuning reshuffled probability $p_{rs}$. For each $p_{rs}$, the reshuffled signed network is obtained and fixed after the reshuffled process where each negative link has the probability $p_{rs}$ to exchange his/her sign with another positive link chosen randomly. The reshuffled signed network is reduced to the real one when $p_{rs}=0$, while all the negative signs will be reshuffled thorough randomly when $p_{rs}=1$. Note that the reshuffled approach provides the possibility and feasibility in comparing the possible signs' organization and the real one. Through analyzing the social status and the opinion formation on all the reshuffled signed networks below, we can reveal the homophily properties of the social status and the dynamic status in the real directed signed social network.

\section{Entropies of social status}

Analogously to an Ising model, a positive link is mapped as one with spin "$\uparrow$" while a negative link mapped as one with spin "$\downarrow$". The presence of negative links introduce disorder (or frustration) in signed social network\cite{facchetti2012}. As well known, information entropy describes the uncertainty associated with a given probability distribution. The application of entropy concept in complex networks is widely and deeply \cite{bianconi2008, anand2011, zhao2011, ye2014, anand2014}. However, the application of entropy in the signed network are presently limited and challenged. Here, The information entropies are calculated to quantify the disorder\cite{timme2014} of sign "+" among out-links and in-links in signed social networks respectively.

\begin{figure}
\begin{center}
\resizebox{0.6\textwidth}{!}{%
  \includegraphics{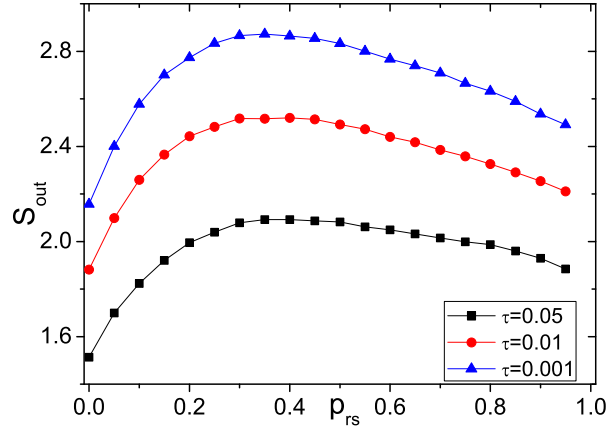}
}
\caption{(Color online)The entropy $S_{out}(p_{rs})$ evolves as a function of the reshuffled probability $p_{rs}$ of the signed social networks of Slashdot081106 with different bin width $\tau$. }
\label{diffbin}
\end{center}
\end{figure}

In signed social network, sign of link describes the social property--such as friend or foe---of the corresponding connection between individuals, while each vertex has his/her social status according to signs of his/her connections\cite{ball2013}. Take vertex $i$ for example, his/her social status can be quantified by a pair of parameters ($p_{i(out)}^{+}, p_{i(in)}^{+}$) to reflect the ratios of positive links among his/her out-links and in-links respectively.

 In detail, the social status\cite{ball2013} for out-links of vertex, says vertex $i$, is defined as follows
\begin{equation}
\label{pioutplus}
 p_{i(out)}^{+}=\frac{k_{i(out)}^{+}}{k_{i(out)}},
\end{equation}
where $k_{i(out)}^{+}$ is the number of the positive links stemmed from $i$, and $k_{i(out)}=k_{i(out)}^{+}+k_{i(out)}^{-}$ denotes $i$'s out-degree, and $p_{i(out)}^{+}$ quantifies the proportion of the positive links directing to his/her local neighbors. Note that $p_{i(out)}^{+}$ is related to his/her local topology structure and the content interacted with his/her local neighbors. Hence, $p_{i(out)}^{+}$ is one of the better physical quantities to quantify $i$'s social status. It is obvious that $0\leq p_{i(out)}^{+}\leq 1$ and the distribution of $p_{i(out)}^{+}$ under the bin width $\tau$ can be written as

\begin{equation}
\label{eq.1}
\pi_{j}^{out}=\frac{\sum\limits_{i=1}^{N_{v}}\delta(j\tau\leq p_{i(out)}^{+}\leq(j+1)\tau)}{N_{v}},
\end{equation}
where $\pi_{j}^{out}$ is the probability that each vertex $i$ with $p_{i(out)}^{+}$ falls in the range of $(j\tau\leq p_{i(out)}^{+}\leq(j+1)\tau)$, and $\delta(x)=1$ when the condition $x$ is true, while $\delta(x)=0$ otherwise. The probability distribution $\{\pi_{j}^{out}|j=0, 1, 2, ..., (\lfloor\frac{1}{\tau}\rfloor-1)\}$ is fixed when $p_{rs}$ is given, i.e., the microscopic state of the system is also fixed.

The entropy of the probability distribution $\{\pi_{j}^{out}|j=0, 1, 2, ..., (\lfloor\frac{1}{\tau}\rfloor-1)\}$ is given by
\begin{equation}
\label{eq.2}
S_{out}(p_{rs})=-\sum_{j=0}^{\lfloor\frac{1}{\tau}\rfloor-1}\pi_{j}^{out}log\pi_{j}^{out},
\end{equation}
which describes the disorder degree of vertices' social status and is called the entropy of social status for out-links. Note that the bin width $\tau$ only change the value of the entropy and does not changes the evolution tendency of the entropy as a function of $p_{rs}$, see Fig.\ref{diffbin}. Hence, we here choose the bin width $\tau = 0.05$.

Comparing the entropy of the system with $p_{rs}=1$ (i.e., all of the negative signs are reshuffled thorough randomly.), the relative entropy of social status for out-links is written as
\begin{equation}
\label{relaentropyout}
\Delta S_{out}(p_{rs})=S_{out}(p_{rs})-S_{out}(p_{rs}=1),
\end{equation}
where $\Delta S_{out}(p_{rs})=0$ means that the signs' organization is thoroughly random, and $\Delta S_{out}(p_{rs})>0$ reflects that the more disorder of signs' organization due to the role of the network topology.

Likewise, we also define the entropy of the probability distribution $\{\pi_{j}^{in}|j=0, 1, 2, ..., (\lfloor\frac{1}{\tau}\rfloor-1)\}$ for in-links
\begin{equation}
\label{eq.3}
S_{in}(p_{rs})=-\sum_{j=0}^{\lfloor\frac{1}{\tau}\rfloor-1}\pi_{j}^{in}log\pi_{j}^{in}
\end{equation}
and the relative entropy of social status for in-links is obtained
\begin{equation}
\label{relaentropyin}
\Delta S_{in}(p_{rs})=S_{in}(p_{rs})-S_{in}(p_{rs}=1),
\end{equation}
where $\pi_{j}^{in}=\frac{\sum\limits_{i=1}^{N_{v}}\delta(j\tau\leq p_{i(in)}^{+}\leq(j+1)\tau)}{N_{v}}$ is the probability that each vertex $i$ with $p_{i(in)}^{+}$ falls in the range of $(j\tau\leq p_{i(in)}^{+}\leq(j+1)\tau)$  and $p_{i(in)}^{+}=\frac{k_{i(in)}^{+}}{k_{i(in)}}$ is the proportion of vertex $i$'s local neighbors who tag $i$ as a friend.

\begin{figure}
\begin{center}
\resizebox{0.6\textwidth}{!}{%
  \includegraphics{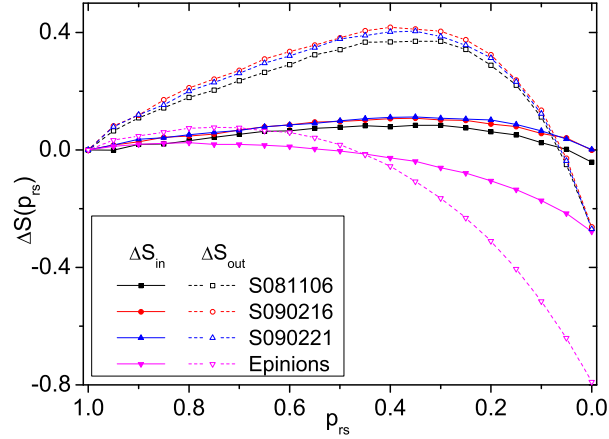}
}
\caption{(Color online)The relative entropy $\Delta S(p_{rs})$ evolves as a function of the reshuffled probability $p_{rs}$. $\Delta S_{in}$ and $\Delta S_{out}$ are the relative entropies of social status for  in-links and out-links respectively.}
\label{entropy}
\end{center}
\end{figure}

The reshuffled signed network is first built using the reshuffled method, i.e., each negative link has the reshuffled probability $p_{rs}$ to exchange its sign with another positive link chosen randomly in the real signed social network. The reshuffled signed network is fixed after the reshuffled process finished. Note that in order to reduce the fluctuation, 50 realizations are made for each given $p_{rs}$.Then the two relative entropies $\Delta S_{out}(p_{rs})$ and $\Delta S_{in}(p_{rs})$ of each shuffled signed social network with $p_{rs}$ are calculated correspondingly. In the special case of $p_{rs}=0$,  $\Delta S_{out}(p_{rs})$ and $\Delta S_{in}(p_{rs})$ are the relative entropies of the real signed social networks. In Fig.\ref{entropy}, we compare the relative entropies of those reshuffled signed networks through analyzing the evolution of  $\Delta S_{out}(p_{rs})$ and $\Delta S_{in}(p_{rs})$ as a function of $p_{rs}$ respectively. The fluctuation of $\Delta S_{in}(p_{rs})$ is smaller than that of $\Delta S_{out}(p_{rs})$ in our several signed social networks, which reflects the stronger of randomness of labeling signs in in-links than that in out-links. For example, $\Delta S_{in}(p_{rs}=0)=0$ shows that the signs of in-link is randomness completely in the Slashdot networks, which shows the irregularities in the sign's organization for in-links of real signed social networks from the global perspective. Furthermore, $\Delta S_{out}(p_{rs})$ reaches its minimum value when $p_{rs}=0$. Namely, although there exists the heterogeneous property of topology structure\cite{ciotti2015}, an increasing number of individuals share the homogeneous property of social status for out-links and in-links in real signed social networks. Last but not least, the surprising result is that the evolution of entropy for out-links in Slashdot networks is different from that in Epinions, which indicates the different topology structures in those two signed social networks. The topology structure enhances the disorder ability of the negative signs in Slashdot networks, i.e., $\Delta S_{out}$ reaches its maximum value when $p_{rs}$ is about $0.35$. However, the topology structure does not play an obviously positive role in the disorder ability of the negative role in Epinions, i.e., $\Delta S_{out}$ increases to its maximum value as $p_{rs}\rightarrow 1$.

\section{Opinion formation on signed social networks}
On the other hand, we focus on the binary-state opinion formation occurring on signed social networks, where each vertex is assumed to hold one of the two possible opinions \{$+1, -1$\} initially random analogous to the spin up and spin down in the Ising model in statistical mechanics. Through analyzing the spreading dynamics of opinion formation, we hope to find the signal of how does dynamics reveal the topology structure of complex networks\cite{timme2007, timme2014}, especially the signs' organization in signed social networks here.

During the interaction between vertices, we consider the initiative of individual to collect information from his/her local neighbors. In our society, individuals around us are diverse and the influence between individuals depends on the properties of their connections, such as the positive links and negative links in signed social networks. For simplicity, the impact factor of vertex $j$ acting on vertex $i$ is quantified as the sign $A_{ij}$ of link stemmed from $i$ to $j$. In other words, $i$ has the same ability to collect information from his/her local neighbors including friends and foes, but the impact factor of his/her friend is reverse to that of his/her foe. ($A_{ij}o_{i}o_{j}$) is the information collected by $i$ from his/her local neighbor $j$. Hence, the information collected by $i$ from his/her local neighbors is quantified\cite{guo2013}
\begin{equation}
\label{eqei}
\Delta E_{i}=\sum_{j\in\Gamma_{i}}A_{ij}o_{i}o_{j},
\end{equation}
where $\Gamma_{i}$ denotes the subgraph composed of vertex $i$ and his/her local neighbors. Eq.(\ref{eqei}) has the same form as the Hamiltonian of the Ising model without the external field in statistical mechanics. Then, we define the flipping probability of vertex $i$'s opinin as a form of Fermi function\cite{grauwin2012, guo2013, liu2014}:,
\begin{equation}
\label{flipping}
f_{i}=[1+exp(\Delta E_{i})]^{-1},
\end{equation}
which reflects the role of human subjective activity in opinion formation and the interaction between individuals in opinion formation.

\begin{figure}
\begin{center}
\resizebox{0.8\textwidth}{!}{%
  \includegraphics{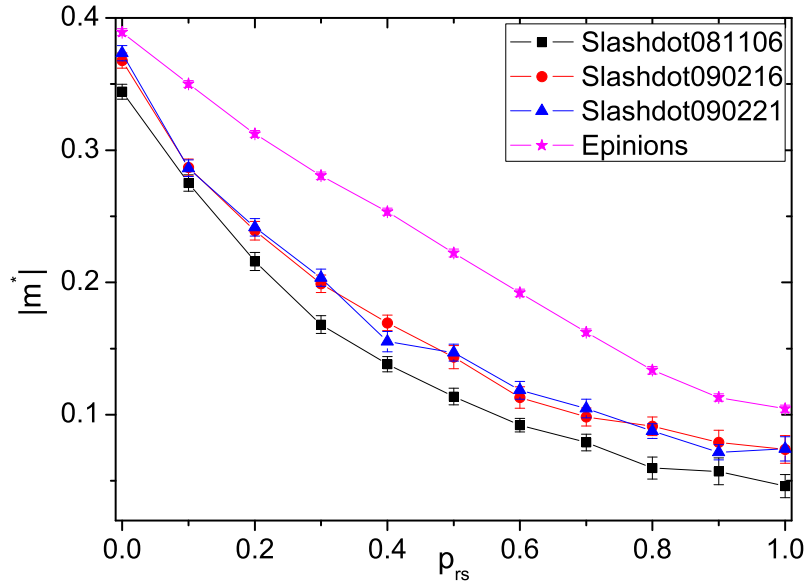}
}
\caption{(Color online)The critical order parameter $|m^{*}|$ of opinion formation varies as the reshuffled probability $p_{rs}$ in the four signed social networks. }
\label{magnet}
\end{center}
\end{figure}

In order to evaluate the real signs' organization in our several signed social networks from the aspect of dynamical process of opinion formation, we realize the majority-rule model of the binary-state opinion formation on each reshuffled network. Firstly, the reshuffled signed network is built using the reshuffled approach and fixed after the reshuffled process finished. Then the opinion formation of the majority-rule model is realized on the fixed reshuffled signed network with $p_{rs}$. At each time step, each vertex collects the information from his/her local neighbors and flips his/her current opinion according to the flipping probability. Next, we calculate the order parameter $m(t)=\frac{1}{N}\sum\limits_{i=1}^{N}o_{j}$, which describes the difference of the number of opinions $+1$ and $-1$ according to the magnetization of the Ising model. The system will reach its stable state as time $t$ elapses\cite{guo2013}, i.e,$\partial m(t)/\partial t|_{t\rightarrow \infty}=0$, and the order parameter $m(t)$ reaches its stable value $|m^{*}|$ when $t \rightarrow \infty $. For each parameter $p_{rs}$, $|m^{*}|$ is calculated and plotted as the function of $p_{rs}$.

 The order state, where all vertices share the same opinion, is also called the consensus state and characterized by $|m^{*}|=1$. Hence, $|m^{*}|$ describes the dynamic status of network from the global perspective. The larger is $|m^{*}|$, the stronger is the consensus ability of the system. That is to say, an increasing number of individuals share the same dynamic status. In Fig.\ref{magnet}, we compare $|m^{*}|$s in different reshuffled signed networks including the real one. Namely, we plot the evolution of $|m^{*}|$ as a function of $p_{rs}$. We find that $|m^{*}|$ decreases as the reshuffled probability $p_{rs}$ increases, and $|m^{*}|$ reaches its maximum value when the reshuffled probability $p_{rs}=0$. The maximum value of $|m^{*}|$ reveals the role of the real signs' organization in promoting the consensus of the opinion formation. Individuals tend to share the homogeneous dynamic status during the collective dynamics of opinion formation which is a common dynamics in society. What's more, the most interesting result is that $|m^{*}|$ increases as the ratio of positive links $p_{+}$ increases at fixed $p_{rs}$ in Fig. \ref{magnet}, which represents the role of negative links in slowing the consensus process of opinion formation and enhancing the diversity in society.

\section{Conclusion and Discussion}
In summary, we have analyzed the signs' organization through calculating the entropies $S_{out}$ and $S_{in}$ of and simulating opinion formation on signed social networks from the global perspective. Firstly, we define the entropies of social status according to the ratio of sign "+" for out-links and in-links and calculate the entropies $S_{out}$ and $S_{in}$. We find that $S_{in}$ is less related to the reshuffled probability $p_{rs}$, which is a powerful evidence for the subjective initiative of individuals in society. While $S_{out}$ reaches its minimum value, which means that the social status of individuals has the homogeneous property although their topology connectivity are heterogeneous. Secondly, we use the collective dynamics of opinion formation to evaluate the signs' organization, and find that the critical order parameter $|m^{*}|$ reaches its maximum value when $p_{rs}\rightarrow 0$. Namely, the real signs' organization enhances the consensus ability of opinion formation in the real directed signed social networks of Epinions and Slashdot. What's more, analogous to the principle of maximum entropy in thermodynamic process, we found that the signs' organization follows the principle of minimum social status entropy. Our present work provides an interesting perspective to understand the direction of sign's organization in directed signed social networks.

\end{document}